\documentclass[12pt,a4paper]{article}

\usepackage{latexsym}
\usepackage{amsmath}
\usepackage{amsfonts}
\usepackage{amssymb}

\newcommand{\qedsymb}{{\em Q.E.D.}}

\newcommand{\be}{\begin{equation}}
\newcommand{\ee}{\end{equation}}
\newcommand{\bea}{\begin{eqnarray}}
\newcommand{\eea}{\end{eqnarray}}

\renewcommand{\theequation}{\arabic{section}.\arabic{equation}}

\def\ad{{\mathrm{ad}}}                  %
\def\G{{\cal G}}                        %
\def\M{{\cal M}}                        %
\def\H{{\cal H}}                        %
\def\A{{\cal A}}                        %
\def\cO{{\cal O}}                       %
\def\ri{{\mathrm{i}}}                   %
\def\rr{{\mathrm{r}}}                   %
\def\cR{{\cal R}}                       %
\def\cL{{\cal L}}                       %
\def\T{{\cal T}}                        %
\def\E{{\cal E}}                        %
\def\bC{{\mathbf C}}                    %
\def\bT{{\mathbf T}}                    %
\def\T{{\cal T}}                        %
\def\dt{\left.{\frac{d}{dt}}\right\vert_{t=0}}      %

\def\bR{{\mathbb R}}                    %
\def\bC{{\mathbb C}}                    %
\def\1{{\mbox{\boldmath $1$}}}          %
\def\tr{\mathrm{tr\,}}                  %
\def\diag{\mathrm{diag}}                %
\def\0{{\mbox{\boldmath $0$}}}          %
\def\cE{{\cal E}}                       %
\def\y{{\cal Y}}                        %

\begin{document}

\vspace*{0.5cm}
\begin{center}
{\Large \bf Spin Calogero models associated with Riemannian symmetric spaces
of negative curvature}
\end{center}

\vspace{0.2cm}

\begin{center}
L. FEH\'ER${}^{a}$ and B.G. PUSZTAI${}^b$ \\

\bigskip

${}^a$Department of Theoretical Physics, MTA  KFKI RMKI\\
1525 Budapest 114, P.O.B. 49,  Hungary, and\\
Department of Theoretical Physics, University of Szeged\\
Tisza Lajos krt 84-86, H-6720 Szeged, Hungary\\
e-mail: lfeher@rmki.kfki.hu

\bigskip
${}^b$Centre de recherches math\'ematiques, Universit\'e de Montr\'eal\\
C.P. 6128, succ. centre ville, Montr\'eal, Qu\'ebec, Canada H3C 3J7, and\\
Department of Mathematics and Statistics, Concordia University\\
7141 Sherbrooke W., Montr\'eal, Qu\'ebec, Canada H4B 1R6\\
e-mail: pusztai@CRM.UMontreal.CA

\end{center}

\vspace{0.2cm}

\begin{abstract}
The Hamiltonian symmetry reduction of the geodesics system
on a
symmetric space of negative curvature by the maximal
compact subgroup of the isometry group is investigated at an
{\em arbitrary} value of the momentum map.
Restricting to regular elements in the configuration space,
the reduction generically yields a spin Calogero
model with hyperbolic interaction potentials defined by
the root system of the symmetric space.
These models come equipped with Lax pairs and many
constants of motion, and can be
integrated by the projection method.
The special values of the momentum map leading to spinless Calogero
models are  classified under some conditions,
explaining why the $BC_n$  models with two
independent coupling constants are
associated with $SU(n+1,n)/S(U(n+1)\times U(n))$
as found by Olshanetsky and Perelomov.
In the zero curvature limit our models reproduce rational spin Calogero models
studied previously  and similar models correspond
to other (affine) symmetric spaces, too.
The construction works at the quantized level as well.

\end{abstract}

\newpage

\section{Introduction}
\setcounter{equation}{0}

The investigation of the structure and applications of
 `Calogero type' models, pioneered in \cite{Cal,Sut,Mos,CRM,Cal2},
is a fascinating subject receiving lots of attention.
It is clear from
the reviews (see~e.g.~\cite{Nekr,Diej,SutWSci}) that
these models appear in extremely many contexts
in physics as well as in mathematics.
The present paper deals with their hyperbolic variants
and extensions
by internal (`spin') degrees of freedom \cite{GH}, at the classical level.
Among alternative approaches to generalized Calogero models,
we are interested in  their relationship
to symmetric spaces,  which was first realized in \cite{OPInv}
and further studied in \cite{OPCim1}-\cite{FW}.

As introduced in \cite{OPInv}, a hyperbolic Calogero type model
is characterized by the Hamiltonian
\be
H(q,p) = \frac{1}{2} \langle p, p\rangle +
\sum_{\alpha \in \cR_+} \frac{g_\alpha^2}{\sinh^2 \alpha(q)},
\label{1.1}\ee
where $\cR_+$ denotes the positive roots in a
root system
$\cR$ and the coupling constants $g_\alpha$ can be different in principle for
different orbits of the corresponding reflection  group.
Here, the crystallographic root systems are considered that occur
in association with symmetric spaces and include,
besides the root systems of the complex simple Lie algebras,
the $BC_n=B_n\cup C_n$ systems \cite{Helg,Knapp}.
If $\cR$ is of the classical $A_n$, $B_n$, $C_n$, $D_n$ or $BC_n$ type,
and the coupling constants are subject to certain relations,
then Olshanetsky and Perelomov were able to construct a Lax representation
and a solution algorithm for the model by treating it as projection of geodesic motion
on a symmetric space of negative curvature \cite{OPInv,OPCim2,OPRep1,Per}.
Their method is equivalent to Hamiltonian symmetry reduction of
the geodesic system by the maximal compact subgroup of the isometry group,
$G_+ \subset G$,
as was explained in the $A_n$ case by Kazhdan, Kostant and Sternberg in \cite{KKS}.
(For general reviews of the theory of Hamitonian reduction, see e.g. \cite{Per,OrRat}.)

The Hamiltonian reduction
yields a Calogero type model (\ref{1.1}) only if
the value of the momentum map
 defining the reduction enjoys some
very specific properties, which are known to occur only for particular symmetric spaces
$G/G_+$, as described in \cite{OPInv,KKS,A,OPRep1,Per,ABT}.
However,  a classification of such `good reductions' is not available.
For reasons not very well understood,
the classical mechanical
 models (\ref{1.1}) based on the exceptional root systems, or on $BC_n$ with
three arbitrary coupling constants, are (up to now) not related to symmetric spaces.
It is also not quite clear why it is the case \cite{OPInv} that
for the $BC_n$ models with two independent coupling constants  the pertinent symmetric space is
$SU(n+1,n)/S(U(n+1) \times U(n))$, although
the root system of
$SU(m,n)/S(U(m)\times U(n))$ is of $BC_n$ type for any $m>n$.
These problems motivated  \cite{dHP,Sas1}  to set  up new frameworks for studying
Calogero models.
Due to its universal applicability, the method developed
in \cite{Sas1,Sas2,Sas3,Sas4} may be considered
more natural than the
traditional Olshanetsky-Perelomov  approach to Calogero type models.
Still, one would like to better understand the relation between these models and
symmetric spaces.

In this paper we reformulate the question about the correspondence between symmetric
spaces of negative curvature and Calogero type models by asking what is the reduced system that results from
the geodesic system  in general, at an {\em arbitrary value} $\mu_0$ of the momentum map
for the action of $G_+$ on $T^*(G/G_+)$.
The answer turns out to be very simple. We demonstrate that the reduction generically yields
a spin Calogero model  with Hamiltonian of the form
\be
H(q,p, \xi) =
\frac{1}{2}\langle p, p \rangle + \frac{1}{2} \sum_{\alpha \in \cR_+}\sum_{i=1}^{\nu_\alpha}
\frac{ \left(\xi^\alpha_i\right)^2}{\sinh^2 \alpha(q)}.
\label{1.2}\ee
The phase space of this model is
$T^* \check \A \times \cO_{red}= \check \A \times \A \times \cO_{red}= \{(q,p,\xi)\}$,
where $\cO_{red}$ is the reduction of the coadjoint orbit of $G_+$ through $-\mu_0$
by the action of a subgroup $M\subset G_+$ at the zero value of its momentum map.
The $\nu_\alpha$ are the multiplicities  of the restricted roots \cite{Helg,Knapp}
with respect to the Cartan subalgebra $\A$ of the symmetric space,
$\check \A$ is the interior of a Weyl chamber and $M$ is the centralizer of $\A$ inside $G_+$.

After deriving the spin Calogero models (\ref{1.2}),
which seem to appear here for the first time,
we show that the Hamiltonian reduction equips them naturally
 with many constants of motion and a spectral parameter dependent Lax pair.
Their evolution equation belongs to a commuting family whose
Hamiltonian flows can be constructed with the aid of the projection method.
The model (\ref{1.2})  simplifies to (\ref{1.1}) if the space of spin degrees of freedom,
$\cO_{red}$,  consists of a single point.
There is only one mechanism known whereby this can be guaranteed.
Namely, if  $G_+$ contains a  simple factor of $SU(k)$ type intersecting $M$
in its maximal torus,
then one can make use of the same orbit of
$SU(k)$ (possibly `dressed' by a contribution from the center of $\mathrm{Lie}(G_+)$),
which was used in \cite{KKS} in relation with the symmetric space $G/G_+=SL(k,{\bC})/SU(k)$.
We shall classify the cases for which this `KKS mechanism' is applicable,
and thereby explain why  the $BC_n$  models with two
independent coupling parameters are
associated with $SU(n+1,n)/S(U(n+1)\times U(n))$
as found by Olshanetsky and Perelomov.

The rational analogues of the models (\ref{1.2}) have  been
obtained recently in \cite{AKLM,Hoch}  by reducing the geodesic motion
on  the symmetric spaces of zero curvature, as initiated in \cite{OPCim1}.
Our results  concerning the list of spinless cases and Lax pairs,
which are not addressed in \cite{AKLM,Hoch}, can  also be applied
in the zero curvature limit.
In \cite{Hoch} the rational spin Calogero models
are presented as an illustration to the general theory of singular
symplectic reduction of cotangent bundles advanced in this paper.
In contrast, we here give a direct, simple
derivation of the models (\ref{1.2}).
We shall proceed similarly to \cite{FP}, where  we obtained
a different class of hyperbolic and trigonometric spin Calogero models
by reducing the geodesic motion on a semisimple Lie group
with the aid of the symmetry induced by twisted conjugations.
Together with the above and several further results  in the literature,
the present work supports the following general statement.
Heuristically formulated, the statement is that
if one reduces geodesic motion on a space of matrices by the Hamiltonian action of a
 symmetry group
whereby those matrices can be diagonalized, then the result
is in general a spin Calogero type model, with coordinate variables parametrizing
the diagonal matrices that arise.
This heuristic statement can be promoted to a proper theorem under various more
precise formulations of the conditions.

The organization of the paper and our results can be outlined as follows.
Section 2 contains necessary background material and conventions.
Our main result is given by Theorem 1 in Section 3 summarizing the outcome
of the derivation of
the reduced Hamiltonian system (\ref{1.2}) from the geodesic motion.
The subsections of Section 4 deal with the conserved quantities
and the Lax representation of this system, with the results formulated in
Theorem 4 and Proposition 5.
Section 5 is devoted to explaining what is meant by the
`KKS mechanism' and to presenting the list of cases in which this mechanism
leads to spinless Calogero models of type (\ref{1.1}).
It is shown that in addition to the original $SL(k,\bC)/SU(k)$ case
the KKS mechanism is applicable only to certain reductions of the
symmetric spaces having $SU(m,n)$ as isometry group for some $m\geq n$,
with the precise list of cases  provided
by Theorem 6.
The corresponding Hamiltonians are collected in Proposition 7,
recovering the classical examples \cite{OPInv,OPRep1,Per}  in our more systematic framework.
Our conclusions are presented in Section 6.
We here briefly discuss also the dynamical $r$-matrices and
the quantization of the models (\ref{1.2}), which will be elaborated in a future publication.
Finally,  Appendix A contains auxiliary material on $su(m,n)$.

\section{Preliminaries on the system to be reduced}
\setcounter{equation}{0}

In this preparatory section we  collect
background material and conventions on Riemannian symmetric spaces of negative curvature.
More details can be found, e.g., in \cite{Per,Helg,Knapp} and
the reader may also consult Section 5
with Appendix A for a concrete example.
Our notations are adapted to matrix Lie groups for simplicity throughout the paper,
but this does  not mean any restriction of generality since
all formulae can be rewritten in a more abstract manner as well.

\subsection{Group theoretic preliminaries and conventions}
\setcounter{equation}{0}

Let $G$ be a non-compact real simple Lie group with finite centre
and $\G$ its Lie algebra. Up to conjugation, there is a unique
Cartan involution $\theta$ of $\G$, which is characterized  by the
decomposition
\be
\G= \G_+ + \G_-, \qquad \theta(X_\pm) = \pm X_\pm
\quad \forall X_\pm \in \G_\pm,
\label{2.1}\ee
where
 the restriction of the Killing form $\langle\ ,\ \rangle$ of $\G$
is negative (resp.~positive) definite on $\G_+$ (resp.~on $\G_-$).
$\G_+$ is a maximal compact subalgebra of $\G$ and the elements of
$\G_-$ are diagonalizable in the adjoint representation of $\G$. Any
maximal Abelian subspace $\A\subset \G_-$ induces the decomposition
\be
\G= \A \oplus \M \oplus \left(\oplus_{\alpha\in \cR}\,
\G_\alpha\right),
\label{2.2}\ee
where
\be
\M:= \{ X\in
\G_+\,\vert\, [H,X]=0\,\,\,\forall H\in \A\,\}, \quad \G_\alpha:= \{
X\in \G\,\vert\, [H,X]=\alpha(H)X\,\,\,\forall H\in \A\,\}.
\label{2.3}\ee
The elements of $\cR\subset \A^*\setminus\{0\}$ are
called restricted roots. We fix a polarization $\cR = \cR_+ \cup
\cR_-$ and choose weight vectors $E_\alpha^i\in \G_\alpha$
($i=1,\ldots, \nu_\alpha:= \mathrm{dim}(\G_\alpha)$) so that
\be
\theta(E_\alpha^i) = - E_{-\alpha}^i, \qquad \langle E_\alpha^i,
E_\beta^j\rangle = \delta_{\alpha, -\beta} \delta^{i,j}.
\label{2.4}\ee
The decomposition (\ref{2.1}) can be refined as
\be
\G_- = \A+ \A^\perp, \qquad \G_+ = \M+ \M^\perp,
\label{2.5}\ee
where $\M^\perp$ and $\A^\perp$ are spanned by the basis vectors
\be
E_\alpha^{+,i} = \frac{1}{\sqrt{2}} ( E_\alpha^i +
\theta(E_\alpha^i)) \in \M^\perp, \qquad E_\alpha^{-,i} =
\frac{1}{\sqrt{2}} ( E_\alpha^i - \theta(E_\alpha^i)) \in \A^\perp
\qquad \forall \alpha\in \cR_+.
\label{2.6}\ee
Lifting $\theta\in \mathrm{Aut}(\G)$ to the
Cartan involution $\Theta$ of $G$, let us introduce
\be
G_+ = \{ g_+
\in G\,\vert\, \Theta(g_+)=g_+\,\}, \qquad G_- = \{ g_- \in
G\,\vert\, \Theta(g_-)= g_-^{-1}\,\}.
\label{2.7}\ee
$G_+$ is  a
maximal compact subgroup of $G$ and the submanifold $G_-\subset G$
is diffeomorphic to $\G_-$ by the exponential map. The group $G$ is
diffeomorphic to $G_- \times G_+$ since any $g\in G$ admits a unique
decomposition as
\be
g= g_- g_+,\quad g_\pm \in G_\pm.
\label{2.8}\ee
The symmetric spaces of negative curvature are the
coset spaces $G/G_+$. A convenient model of such a coset space is
provided by the identification
\be
G/G_+ \simeq G_-,
\label{2.9}\ee
where the corresponding projection $\pi: G\to G_-$ is by definition
given by
\be
\pi: g\mapsto \Lambda(g):= g \Theta(g^{-1}) = g_-^2
\quad \hbox{for}\quad g=g_- g_+.
\label{2.10}\ee
The left
translation on $G$ by $\eta\in G$ descends to the action on the
symmetric space $G/G_+$, which operates according to
\be
G\ni\eta \mapsto \rho_\eta \in
\mathrm{Diff}(G/G_+), \qquad \rho_\eta(\Lambda) = \eta \Lambda
\Theta(\eta^{-1}).
\label{2.11}\ee

\subsection{Hamiltonian model of the geodesic motion on $G/G_+$}

Later we shall reduce the Hamiltonian system of the geodesic motion on $G/G_+$
using the action of the symmetry group $G_+$ induced by (\ref{2.11}).
A very convenient model of this Hamiltonian system can be
obtained by reducing the geodesic system on $T^*G$ by the $G_+$ action
defined by right translations, fixing the corresponding momentum map to zero.
Indeed, as is easily verified, this leads to the model
\be
(T^*(G/G_+), \Omega, \H),
\label{2.12}\ee
where the various ingredients are identified as follows.
First, the phase space is
\be
T^*(G/G_+) \simeq T^* G_- \simeq G_- \times \G_-
=\{ (\Lambda, J_-)\,\vert\, \Lambda \in G_-,\,\, J_-\in \G_-\,\}.
\label{2.13}\ee
To describe the symplectic form $\Omega$ and the Hamiltonian $\H$,
let us introduce the $\G_+$ valued function $J_+$ by the formula
\be
J_+(\Lambda, J_-) = (\tanh \ad_Q) J_-\quad\hbox{with}\quad
Q:= \frac{1}{2} \log \Lambda,
\label{2.14}\ee
which is well-defined since $\ad_Q$ has real eigenvalues only.
Then introduce $J: T^*(G/G_+) \to \G$ by
\be
J(\Lambda, J_-) = J_- + J_+(\Lambda, J_-).
\label{2.15}\ee
Note  that the defining equation of $J_+$ can be
rewritten as
\be
\Lambda^{-1} J \Lambda = J_- - J_+.
\label{2.16}\ee
Now the symplectic form and the geodesic Hamiltonian are
\be
\Omega= d\vartheta
\quad\hbox{with}\quad
\vartheta = \frac{1}{2} \langle J, d\Lambda \Lambda^{-1}\rangle,
\qquad
\H= \frac{1}{2} \langle J, J\rangle.
\label{2.17}\ee
The Hamiltonian action of $G$ operates on the phase space (2.13) by
$\rho_\eta^*\in  \mathrm{Diff}(T^*G_-)$,
\be
\rho^*_\eta: (\Lambda, J_-) \mapsto
 (\eta \Lambda \Theta(\eta^{-1}), (\eta J(\Lambda, J_-) \eta^{-1})_-),
 \qquad \forall \eta\in G,
\label{2.18}\ee
where we use the decomposition $X=X_++ X_-$ for any $X\in \G$.
In fact,  $J: T^*(G/G_+) \to \G$ is nothing but the equivariant momentum map
that generates this action.
This means that if $T_a$ is a basis of $\G$, and
\be
J_a = \langle T_a, J \rangle,
\qquad
[T_a, T_b] = f_{ab}^c T_c,
\label{2.19}\ee
then we have the Poisson brackets
\be
\{ \Lambda, J_a\} = T_a \Lambda - \Lambda \theta(T_a),
\qquad
\{ J_a, J_b\} = f_{ab}^c J_c.
\label{2.20}\ee
Naturally, $J_+: T^*(G/G_+) \to \G_+$ is the momentum map for the
restriction of the above action to $G_+$, which simplifies according to
\be
\rho^*_\eta: (\Lambda, J_-) \mapsto (\eta \Lambda \eta^{-1}, \eta J_- \eta^{-1})
\qquad
\forall \eta \in G_+.
\label{2.21}\ee

The Hamiltonian equations of motion can be written as
\be
\dot{\Lambda} = \{ \Lambda, \H\} = J \Lambda - \Lambda \theta(J),
\qquad
\dot J_- = \{ J_-, \H\} =0.
\label{2.22}\ee
By (\ref{2.16}) the first formula is equivalent to
\be
\dot{\Lambda}\Lambda^{-1}+ \Lambda^{-1} \dot{\Lambda}= 4 J_-.
\label{2.23}\ee
The solution with initial value $(\Lambda_0, J_-^0)$ is just
the orbit of the one-parameter subgroup of $G$ generated by
$J_0 := J(\Lambda_0, J_-^0)$:
\be
\Lambda(t) = e^{t J_0} \Lambda_0 e^{-t \theta(J_0)},
\label{2.24}\ee
and the components of $J$ are constants of motion.

\section{Spin Calogero models from Hamiltonian reduction}
\setcounter{equation}{0}

This section contains our derivation of spin Calogero models from the geodesic motion
on the symmetric space, with the result  given by Theorem 1 and subsequent remarks.

We below use the subset
of regular elements $\hat \A\subset \A$,
\be
\hat \A = \{ H\in \A\,\vert\, \alpha(H)\neq 0\quad \forall \alpha\in \cR\,\},
\label{3.1}\ee
and the open Weyl chamber
\be
\check \A:= \{ H\in \A\,\vert\, \alpha(H)>0 \quad \forall \alpha\in \cR_+\,\},
\label{3.2}\ee
which is a connected component of $\hat \A$.
The $G_+$-conjugates of $\check \A$ form a dense open subset
$\check \G_-\subset \G_-$, and we focus on
the corresponding dense open submanifold of
 $T^*(G/G_+)$ furnished by
\be
P:= T^* \check G_-,
\qquad
\check G_- := \exp(\check\G_-).
\label{3.3}\ee
We wish to reduce $P$ under the Hamiltonian action of $G_+$ at an {\em arbitrary} value,
$\mu_0$, of the momentum map $J_+$.
To characterize the Marsden-Weinstein reduction of the Hamiltonian system
\be
(P, \Omega, \H),
\label{3.4}\ee
we make use of the standard shifting trick of symplectic reduction (see, e.g., \cite{OrRat}).
For this, we let
\be
(\cO, \omega^\cO)
\label{3.5}\ee
denote the coadjoint orbit of $G_+$ through $(-\mu_0)$ equipped with its
natural symplectic form $\omega^\cO$,
where $\G_+^*$ is identified with $\G_+$ by the Killing form.
The shifting trick states that the reduced system mentioned above is
naturally isomorphic to the Marsden-Weinstein reduction of the `extended system'
\be
(P^\cO, \Omega^\cO, \H^\cO)
\label{3.6}\ee
at the {\em zero} value of the appropriate momentum map, $\Psi$.
The extended system is
\be
P^\cO= P\times \cO,
\qquad
\Omega^\cO = \Omega + \omega^{\cO},
\qquad
\H^\cO(\Lambda, J_-, \xi) = \H(\Lambda, J_-),
\label{3.7}\ee
where $(\Lambda, J_-, \xi)\in P^\cO$ is arbitrary.
Using (\ref{2.15}),  $\H^\cO$ can  be written as
\be
\H^\cO=\frac{1}{2}\langle J^\cO, J^\cO\rangle
\quad\hbox{with}\quad
J^\cO(\Lambda, J_-, \xi) := J(\Lambda, J_-).
\label{extra1}\ee
The action of $G_+$ on $P^\cO$ is the diagonal one, denoted as $\hat\rho$:
\be
\hat\rho_\eta:    (\Lambda, J_-, \xi) \mapsto
(\eta \Lambda \eta^{-1}, \eta J_- \eta^{-1},
\eta \xi \eta^{-1}),
\qquad
\forall \eta \in G_+,
\label{3.8}\ee
and this is generated by the momentum map
\be
\Psi: P^\cO \to \G_+, \qquad
\Psi(\Lambda, J_-, \xi) = J_+(\Lambda, J_-) + \xi.
\label{3.9}\ee
With $G_+(\mu_0)$ being the isotropy group of $\mu_0$,
the main point is the second equality in
\be
P_{red}:= P_{J_+=\mu_0}/G_+(\mu_0) = P^{\cO}_{\Psi=0}/G_+.
\label{3.10}\ee

After the foregoing preparations, we are now in the position to describe the
reduced Hamiltonian system.
The crucial step is to observe that all $G_+$ orbits in the constrained manifold
$P^\cO_{\Psi=0}$  intersect the following  gauge slice:
\be
S:= \{ (e^{2q}, J_-, \xi)\in P^\cO_{\Psi=0}\,\vert\, q\in \check \A\,\},
\label{3.11}\ee
since every regular element of $\G_-$ can be conjugated into $\check \A$.
This gauge slice is `thick' in the sense that it represents only a partial
gauge fixing of the `gauge transformations' defined by the $G_+$ action.
In fact, the residual gauge transformations (that map
an arbitrarily chosen point of $S$ into $S$) are generated
precisely by the centralizer subgroup $M$ of $\A$ inside $G_+$:
\be
M:=\{ m\in G_+\,\vert\, m H m^{-1} =H \quad \forall H\in \A\}.
\label{3.12}\ee
Therefore we obtain the identification
\be
P_{red}=P^{\cO}_{\Psi=0}/G_+ = S/M.
\label{3.13}\ee
To proceed further, let us decompose $J_-\in \G_-$ and $\xi\in \cO\subset\G_+$
according to (\ref{2.5}) as
\be
J_-= J_\A + J_{\A^\perp},
\qquad
\xi=\xi_\M + \xi_{\M^\perp}.
\label{3.14}\ee
Then one can check that the constraint $\Psi=0$ on $S$
is equivalent to the requirements
\be
\xi_\M=0
\qquad\hbox{and}\qquad
J_{\A^\perp} = - (\coth \ad_q) \xi_{\M^\perp}.
\label{3.15}\ee
This motivates to consider the smooth one-to-one map
\be
I: (\check \A \times \A) \times (\cO\cap \M^\perp) \to S,
\qquad
I(q,p, \xi_{\M^\perp}):=(e^{2q}, p - (\coth \ad_q) \xi_{\M^\perp}, \xi_{\M^\perp}).
\label{3.16}\ee
The pull-back of
$\Omega^{\cO}\vert_S$ by $I$ turns out to be
\be
I^* (\Omega^{\cO}\vert_S)= d \langle p, dq\rangle + \omega^{\cO}\vert_{\cO\cap \M^\perp}.
\label{3.17}\ee
The first term is the canonical symplectic structure of
\be
T^* \check \A = \check \A \times \A = \{ (q,p)\}.
\label{3.18}\ee
The second term in (\ref{3.17}) is the restriction of $\omega^\cO$
to the zero level set of the momentum map for the action of the subgroup $M\subset G_+$ on $\cO$,
which is provided by $\cO\ni \xi \mapsto \xi_\M \in \M\simeq \M^*$.
It is also important to note that $I$ is an $M$ equivariant map,
where $M$ acts trivially on $T^*\check \A$.
As for the reduced Hamiltonian of the geodesic motion,
we  find from (\ref{extra1})
\be
(\H^\cO \circ I)(q,p, \xi_{\M^\perp} ) = \frac{1}{2} \langle L(q,p,\xi_{\M^\perp}),
L(q,p,\xi_{\M^\perp})\rangle
\label{3.19}\ee
with the map $L:= J^\cO\circ I: T^* \check \A \times (\cO\cap \M^\perp) \to \G$,
which is equivariant under the natural actions of $M\subset G_+\subset G$.
By expanding $\xi_{\M^\perp}$ in the basis (\ref{2.6}),
\be
\xi_{\M^\perp}= \sum_{\alpha\in \cR_+}\sum_{i=1}^{\nu_\alpha}\xi^\alpha_i E_\alpha^{+,i},
\label{3.20}\ee
$L$ can be written explicitly as
\be
L(q,p,\xi_{\M^\perp})  = p -
(\coth \ad_q) \xi_{\M^\perp} - \xi_{\M^\perp} = p -\sum_{\alpha\in \cR_+}\sum_{i=1}^{\nu_\alpha}
\xi^\alpha_i \left( \coth\alpha(q) E_\alpha^{-,i} + E_\alpha^{+,i}\right).
\label{3.21}\ee
On account of its equivariance property,
the map $I$ (\ref{3.16}) gives rise to the identification
\be
S/M= T^*\check \A \times (\cO \cap \M^\perp)/M.
\label{3.22}\ee
Combining this with (\ref{3.13}) proves the following theorem.

\medskip
\noindent
{\bf Theorem 1.}
{\em
The reduction of the geodesic system on $\check G_-\subset G/G_+$
defined by (\ref{3.10}) with (\ref{3.3})
can be identified as
$(P_{red}, \Omega_{red}, \H_{red})$  with
\be
P_{red} = T^* \check \A \times \cO_{red},
\qquad
\Omega_{red}= d \langle p, dq\rangle + \omega_{red}^\cO,
\label{3.23}\ee
where $q,p$ are the natural variables on $T^*\check \A$ and
$(\cO_{red}, \omega^\cO_{red})$
is the  symplectic reduction of $(\cO, \omega^\cO)$
by the subgroup $M\subset G_+$ (\ref{3.12}) at the zero value of its momentum map,
\be
\cO_{red}= (\cO\cap \M^\perp)/M .
\label{3.24}\ee
The reduced Hamiltonian defines a hyperbolic spin Calogero type model in general,
since as an $M$ invariant function
on $T^* \check \A \times \cO\cap \M^\perp$ it has the form
\be
\H_{red}(q,p, \xi_{\M^\perp})= \frac{1}{2}
\langle L(q,p,\xi_{\M^\perp}), L(q,p,\xi_{\M^\perp})\rangle
= \frac{1}{2}\langle p, p \rangle + \frac{1}{2} \sum_{\alpha \in \cR_+}\sum_{i=1}^{\nu_\alpha}
\frac{ \left(\xi^\alpha_i\right)^2}{\sinh^2 \alpha(q)}.
\label{3.25}\ee
}

\medskip\noindent
{\bf Remark 2.}
Instead of $S$ (\ref{3.11}), one could equally well use the slightly `thicker'
 gauge slice
\be
\hat S:= \{ (e^{2q}, J_-, \xi)\in P^\cO_{\Psi=0}\,\vert\, q\in \hat{\A}\,\},
\label{3.26}\ee
where $\hat \A$ (\ref{3.1}) is the union of all open Weyl chambers.
The residual gauge transformations now belong to the normalizer
\be
\hat M:=\{ n\in G_+\,\vert\, n H n^{-1} \in \A \quad \forall H\in \A\}.
\label{3.27}\ee
Recalling that $M \subset \hat M$ is a normal subgroup and
\be
W:= \hat M/M
\label{3.28}\ee
is the Weyl group of the symmetric space, we obtain
\be
P_{red}= \hat S/\hat M =(\hat S/M)/(\hat M/M) = \hat P_{red}/W
\ee
with
\be
\hat P_{red}:= \hat S/M = T^* \hat \A \times \cO_{red}.
\label{3.29}\ee
Here,  $\hat P_{red}$ differs from  $P_{red}$
only in that $q$ now varies in $\hat \A$.
The geodesic system descends to a spin Calogero type  system on $\hat P_{red}$,
with Hamiltonian still of the form (\ref{3.25}).
This system
enjoys Weyl symmetry, where $W$  acts on all three components of
$(q,p, [\xi_{\M^\perp}])\in \hat P_{red}$ naturally.
All our spin Calogero models possess a hidden Weyl group symmetry in this sense.

\medskip\noindent
{\bf Remark 3.}
The reduced phase space $P_{red}$ (\ref{3.23}) is not a smooth manifold in general, since the
space of `spin' degrees of freedom has some singularities.
For example, if $\G$ is a real split simple Lie
algebra (like $sl(n,\bR)$), then $\M=\{0\}$,
$M$ is a finite group
and $P_{red}= T^*\check \A \times (\cO/M)$ is an orbifold.
In general, $\cO_{red}$ (\ref{3.24}) is a
stratified space, whose strata are smooth symplectic manifolds \cite{OrRat}.
The restriction to the principal orbit type for
the $M$-action on $\cO\cap \M^\perp$ always leads to a dense open subset of $\cO_{red}$,
which is a  smooth manifold.
A detailed study of the non-principal strata appears as an interesting problem for the future.
In certain  special cases
it so happens that $\cO_{red}$ is a trivial manifold
consisting of a single point, and then the reduced system is a Calogero type model (\ref{1.1})
without spin.
This is further discussed in Section 5.

\section{Constants of motion and Lax pairs}
\setcounter{equation}{0}

The $G_+$ invariant constants of motion of the extended system
(\ref{3.6})  survive the Hamitonian reduction to $P_{red}$ (\ref{3.10}).
By using this we exhibit a large family of conserved quantities
for the spin Calogero model of Theorem 1,
and prove that those of them that are associated (by equation (\ref{4.5})) with the
$G$ invariant functions on $\G$ are in involution.
Then we show that these conserved quantities
in involution admit the usual interpretation as
$G$ invariant functions of a suitable (spectral
parameter dependent) Lax operator for the spin Calogero model.
The commuting constants of motion include the
$G$ invariant functions of the symmetry generator $J$ (\ref{2.15}),
for which the reduced Hamiltonian flows are easily obtained by the
projection method.

\subsection{Constants of motion}

Observe from (\ref{3.7}) that
 $J_-$ and $\xi$ are conserved quantities for the system (\ref{3.6}).
 Therefore so is their  linear combination $K(x): P^\cO \to \G$ given by
 \be
 K(x):=J_- - x\, \xi,
\label{4.1}\ee
 where $x$ is an arbitrary real number.
 In the definition of $K(x)$ we regard $J_-$ and $\xi$
 as evaluation functions on the phase space, i.e.,
 $K(x): P^\cO\ni (\Lambda, J_-,\xi)\mapsto J_- - x\xi \in \G$.
 Since $K(x)$ is equivariant with respect to the natural actions
 of the symmetry group $G_+$ on
 $P^\cO$ and on $\G$,
 the composite
$f\circ K(x)$ is a $G_+$ invariant function on $P^\cO$ for any
 $G_+$ invariant (real) function on $\G$, $f\in C^\infty_{G_+}(\G)$.
Here and below we use the notations
\begin{eqnarray}
&& C^\infty_{G_+}(\G):=\{f\in C^\infty(\G)\:|\: f(gXg^{-1})=f(X)\quad
\forall X\in\G,\forall g\in G_+\},\qquad
\label{4.2}\\
&& C^\infty_{G}(\G):=\{f\in C^\infty(\G)\:|\: f(gXg^{-1})=f(X)\quad
\forall X\in\G,\forall g\in G\}.
\label{4.3}\end{eqnarray}

Any $G_+$ invariant (smooth) function  on $P^\cO_{\Psi=0}$ can be regarded as a (smooth)
function on the reduced phase space $P_{red}$  defined by (\ref{3.10}).
In particular, if
\be
{\cal E}: P^\cO_{\Psi=0} \to P^\cO
\label{4.4}\ee
is the tautological embedding, then
\be
f\circ K(x)\circ \cE \in C^\infty(P_{red})
\qquad
\forall f\in C^\infty_{G_+}(\G).
\label{4.5}\ee
\emph{All functions of this form  are constants of motion
for the reduced system of Theorem 1.}
The Poisson brackets of these functions under the reduced Poisson structure
on $P_{red}$ are given by
\be
\{ f\circ K(x)\circ \cE , h\circ K(y)\circ \E\}_{red} :=
\{ f\circ K(x), h\circ K(y)\}\circ \cE,
\quad \forall f,h\in C^\infty_{G_+}(\G),\,\, \forall x,y\in \bR.
\label{4.6}\ee
On the right-hand-side the Poisson bracket of $P^\cO$ is used,
whose explicit form is determined by (\ref{2.20}) together with the $\G_+$ Lie-Poisson brackets of
the components of $\xi$.

For any real function $f\in C^\infty(\G)$, its gradient
$\nabla f\in C^\infty(\G,\G)$ is defined by
\begin{equation}
\dt f(X+tY)=\langle Y,(\nabla f)(X)\rangle,\quad
\forall X,Y\in\G,
\label{4.7}
\end{equation}
and, using also (\ref{2.1}),
the infinitesimal versions of the invariance conditions (\ref{4.2}), (\ref{4.3})  read
\begin{eqnarray}
&& [X,(\nabla f)(X)]_+ = 0,\quad\forall f\in C^\infty_{G_+}(\G),\, X\in \G,
\label{4.8}\\
&& [X,(\nabla f)(X)] = 0,\quad\forall f\in C^\infty_G(\G),\, X\in \G.
\label{4.9}\end{eqnarray}
To formulate our next result,  we again refer to (2.1) and introduce the decomposition
\be
\nabla f = (\nabla f)_+ + (\nabla f)_-,
\qquad
(\nabla f)_\pm \in C^\infty(\G,\G_\pm).
\label{4.10}\ee

\medskip
\noindent
{\bf Theorem 4.}
\emph{The constants of motion of the spin Calogero model of Theorem 1 that are
provided by equation (\ref{4.5}) satisfy the Poisson bracket relation
\begin{eqnarray}
\{ f\circ K(x),h\circ K(y) \} \circ \cE = xy\langle\xi ,\left[(\nabla f)_+\circ K(x),
(\nabla h)_+\circ K(y) \right]\rangle \circ \E && \nonumber\\
-\langle\xi ,\left[(\nabla f)_-\circ K(x),
(\nabla h)_-\circ K(y)\right]\rangle \circ \E &&
\label{4.11}\end{eqnarray}
$\forall$ $f,h\in C^\infty_{G_+}(\G)$ and $x,y,\in \bR$, with $\xi$ being the $\cO$ valued
 evaluation function on $P^\cO$ (\ref{3.7}).
This Poisson bracket vanishes identically
for any $x$ and $y$ if both $f$ and $h$ belong to $C^\infty_G(\G)$.
It also vanishes identically
$\forall f\in C^\infty_{G_+}(\G)$, $x\in \bR$ if
 $h\in C^\infty_{G}(\G)$ and $y^2=1$.
}

\medskip
\noindent
{\bf Proof.}
Formula (\ref{4.11}) itself is readily calculated by using the Poisson bracket on
$P^\cO$ and imposing the constraint $\Psi = J_+ + \xi =0$ at the end of the calculation.
To verify the claimed involution properties,
we introduce the shorthand
\be
A^f(x):= A_+^f(x)+A_-^f(x):= (\nabla f)\circ K(x)
\label{4.12}\ee
with the subscripts referring to (\ref{2.1}).
Now, for any $f\in C^\infty_{G_+}(\G)$ and $h\in C^\infty_G(\G)$,
notice that the identity
\be
 x\langle\xi,[A^f_+(x),A^h_+(y)]\rangle = y\langle\xi,[A^f_-(x),A^h_-(y)]\rangle
\label{4.13}\ee
is valid for all $x,y\in\mathbb{R}$. Indeed,
this  comes from the following calculation
\begin{eqnarray}
\lefteqn{x\langle\xi,[A^f_+(x),A^h_+(y)]\rangle=}\nonumber\\
&& =-\langle K(x),[A^f_+(x),A^h_+(y)]\rangle
=-\langle K(x),[A^f(x),A^h_+(y)]\rangle +\langle K(x),[A^f_-(x),A^h_+(y)]\rangle\nonumber\\
&& =\langle K(y),[A^f_-(x),A^h_+(y)]\rangle
=\langle K(y),[A^f_-(x),A^h(y)]\rangle-\langle K(y),[A^f_-(x),A^h_-(y)]\rangle\nonumber\\
&& =y\langle\xi,[A^f_-(x),A^h_-(y)]\rangle.
\label{4.14}\end{eqnarray}
By applying this identity, (\ref{4.11}) gives
\begin{equation}
\{f\circ K(x) ,h\circ K(y)\}\circ \cE
=(y^2-1)\langle\xi,[A_-^f(x),A^h_-(y)]\rangle \circ \cE,
\label{4.15}\end{equation}
which implies the last sentence of the theorem.
If both $f$ and $h$ belong to $C^\infty_G(\G)$, then similarly to (\ref{4.13})  we obtain
\be
 y\langle\xi,[A^f_+(x),A^h_+(y)]\rangle = x\langle\xi,[A^f_-(x),A^h_-(y)]\rangle.
\label{4.16}\ee
By combining (\ref{4.13}) and (\ref{4.16}),
it follows that
\begin{equation}
(x^2-y^2)\langle\xi,[A^f_-(x),A^h_-(y)]\rangle \equiv 0,
\label{4.17}\end{equation}
and  by introducing the open planar subset
$\mathcal{D}:=\mathbb{R}^2\setminus\{(x,y)\:|\: x=\pm y\}$, we
see that
\begin{equation}
\langle\xi,[A_-^f(x),A_-^h(y)]\rangle\equiv 0,\quad\forall (x,y)\in\mathcal{D}.
\label{4.18}\end{equation}
Thus (\ref{4.15}) implies
\begin{equation}
\{f\circ K(x) ,h\circ K(y)\}(m)=0\quad\forall (x,y)\in\mathcal{D},\,
\forall m\in P^\cO_{\Psi=0}.
\label{4.19}\end{equation}
Since the function
$\mathbb{R}^2\ni(x,y)\mapsto\{f\circ K(x),h\circ K(y)\}(m)\in\mathbb{R}$
is continuous, it is necessarily zero on the closure of $\mathcal{D}$.
This proves that (\ref{4.11}) vanishes indeed for
all $x,y\in\mathbb{R}$ if $f$ and $h$ are $G$ invariant functions on $\G$. \qedsymb

\medskip

Tracing  the definitions, one sees that the spin Calogero Hamiltonian
$\H_{red}$ (\ref{3.25}) can be identified as
\be
\H_{red}= h_2 \circ K(\pm 1) \circ \cE
\quad
\hbox{with}\quad
h_2(X) = \frac{1}{2}\langle X, X\rangle \quad \forall X\in \G,
\label{4.20}\ee
and $h_2 \circ K(x) \circ \cE$ for any $x$ differs from $\H_{red}$ only
by a multiple of the irrelevant
Casimir function $\langle \xi,\xi\rangle$.
Taking arbitrary $f\in C^\infty_G(\G)$ and $x\in \bR$, (\ref{4.5}) yields a family
of functions in involution that contain the spin Calogero Hamiltonian.
This could be  sufficient for the Liouville integrability of the reduced system
on a generic (or any) symplectic leaf, but counting
the number of independent invariants is tricky and we do not deal with it here.

Now we explain how the Hamiltonian flows of $\H_{red}$ and its constants of
motion in involution considered below can be determined by
the projection method.
We start by observing that the functions $K(1)$ defined in (\ref{4.1})  and $J^\cO$
defined in (\ref{extra1})
coincide on the constrained manifold $P^\cO_{\Psi=0}$.
Consequently, we have
\be
f\circ K(1)\circ \cE= f \circ J^\cO \circ \cE
\qquad
\forall f\in C^\infty_G(\G).
\label{4.21}\ee
This means that the functions $f\circ K(1)$ and $f \circ J^\cO$ are the same from
the point of view of the reduced system, whence
their reduced Hamiltonian flows are also the same.
The Hamiltonian flow of $f\circ J^\cO \in C^\infty(P^\cO)$ with
any initial value $(\Lambda_0, J_-^0, \xi^0)\in P^\cO$ is given explicitly by
\be
(\Lambda(t), J_-(t), \xi(t))=
(e^{t \nabla f (J_0)} \Lambda_0 e^{-t \theta( \nabla f(J_0))},
J_-^0, \xi^0),
\qquad
J_0 := J_-^0 + J_+(\Lambda_0, J_-^0).
\label{4.22}\ee
The flow (\ref{4.22}) preserves $P^\cO_{\Psi=0}$ and its
projection to the reduced phase space integrates the Hamiltonian
vector field of the function (\ref{4.21}) regarded as an
element  of $C^\infty(P_{red})$.
Developed in more detail,
one can find the flows induced on $P_{red}$  by
the conserved quantities (\ref{4.21}) as follows.
The first step is to determine
$\Lambda_0 = e^{2q_0}$ and $J_-^0= p_0 - (\coth \ad_{q_0})\xi^0$
from the initial value $(q_0, p_0, [\xi^0]) \in P_{red}$,
where $\xi^0$ is any representative of $[\xi^0]\in \cO_{red}$.
The second step is to find the curve (\ref{4.22}).
Finally, one projects this curve to the reduced phase space by
 diagonalizing  $\Lambda(t)$ as
$\Lambda(t) = g_+(t) e^{2q(t)} g_+^{-1}(t)$ with $g_+(t)\in G_+$,
whereby  $q(t)$ gives the trajectory in $\check \A$, at least for small $t$.
(It can in principle occur that $q(t)$ reaches the boundary of $\check \A$
at finite $t$, which corresponds to the incompleteness of the
Hamiltonian vector field on $P_{red}$.)
Incidentally, the
set of  functions (\ref{4.21}) coincides
with $\{ f \circ K(-1) \circ \cE\,\vert\, f\in C^\infty_G(\G)\}$.
For different conserved quantities, if exist,
one must use a more complicated algorithm to find the flows.

In view  of the involution properties given by Theorem 4, one may wonder
if there exist any $G_+$ invariant functions for
which (\ref{4.11}) is non-vanishing, for some orbit $\cO$ of some group $G$.
We shall furnish examples of such functions at the end of Section 5.

\subsection{Lax representation}

In order to find a Lax pair for the
system given by Theorem 1, let us start with a remark
on how to obtain the Hamiltonian vector field of the reduced system
in correspondence with an invariant Hamiltonian in general.
Namely, suppose that $V$ is the Hamiltonian vector field
before reduction
and $\sigma$ is the gauge slice of a (partial or complete) gauge fixing
in the constrained manifold defined by the momentum map constraint.
Then the reduced evolution equation is generated by a vector field $V^*$ on $\sigma$,
which always has the form
\be
V^* = V\vert_\sigma  + Y,
\label{4.23}\ee
where $Y$ is the generator of certain infinitesimal gauge transformations.
The `correction term' $Y$ is (partially or completely) determined by the condition that
$V^*$ must be tangent to $\sigma$.
In the case of a complete gauge fixing, $V^*$ is the Hamiltonian vector field with
respect to the reduced Poisson bracket (alias the Dirac bracket)
associated with the gauge slice $\sigma$.

We now take $\sigma$ to be either the `thick slice' $S$ (\ref{3.11}) or
the cross section of a (local) complete gauge fixing inside $S$.
We can parametrize the general element of $\sigma$ as a triple
\be
(e^{2q}, L_-, \xi_{\sigma})
\quad\hbox{with}\quad
L_- = p - \coth(\ad_q) \xi_{\sigma},
\label{4.24}\ee
where $q\in \check \A$, $p\in \A$ and $\xi_{\sigma}\in \cO\cap \M^\perp$
with further restrictions on the form of  $\xi_{\sigma}$
if $\sigma$ is a complete gauge fixing.
The derivatives, $\cL_{V^*}$, of these variables along $V^*$ are subject to
\be
 \cL_{V^*} (q) =  L_- - \frac{1}{2} \sinh (2\ad_q) \y,
 \quad
 \cL_{V^*} (L_-) = [\y, L_-],
 \qquad
\cL_{V^*} (\xi_{\sigma}) = [ \y, \xi_{\sigma}],
\label{4.25}\ee
where $\y$ is a $\G_+$-valued function on $\sigma$ realizing the term
$Y$ in (\ref{4.23}).
The formulae in (\ref{4.25}) follow by combining the Hamiltonian vector field $V$  of
the system (\ref{3.6}), which can be read off from (\ref{2.22}) and
$\cL_V (\xi) =0$, and the infinitesimal variant of the gauge transformations (\ref{3.8}).
By decomposing the gauge transformation parameter $\y$ using (\ref{2.5}),
\be
\y = \y_\M + \y_{\M^\perp},
\label{4.26}\ee
the relation $\cL_{V^*} (q)\in \A$ and the form of $L_-$
(\ref{4.24}) fix $\y_{\M^\perp}$ uniquely as
\be
\y_{\M^\perp }= - w^2(\ad_q) \xi_{\sigma}
\label{4.27}\ee
with the analytic function
\be
w(z)= (\sinh z )^{-1}.
\label{4.28}\ee
The component $\y_\M$ is arbitrary if $\sigma =S$ (when it gives
a residual gauge transformation), and
in general it is subject to the requirement that the form of
$\cL_{V^*} (\xi_{\sigma})$ must be consistent with
the gauge fixing conditions imposed on $\xi_\sigma$.
By (\ref{4.25}) and (\ref{4.27}), $\y_\M$ can be taken to be independent
of $p$, but depends on $q$ and $\xi_\sigma$ in general.
Identifying $\cL_{V^*}$ with the evolutional
derivative, denoted by dot,
we immediately obtain the following result.

\medskip
\noindent
{\bf Proposition 5.}
{\em
Let us describe the spin Calogero system of Theorem 1 using a gauge slice
$\sigma \subseteq S$ parametrized by (\ref{4.24}).
Then the evolution equation can be written as
$\dot q=p$ and
\bea
&&\dot p = [ w^2(\ad_q)\xi_\sigma, \coth (\ad_q) \xi_\sigma]_\A,
\label{4.29} \\
&&\dot \xi_\sigma =[ \y_\M - w^2(\ad_q) \xi_\sigma, \xi_\sigma],
\label{4.30}\eea
where
$\y_\M: \sigma \to \M$ yields an infinitesimal gauge transformation
so that (\ref{4.30}) is consistent with the form of $\xi_\sigma$.
Defining the functions $L(x):\sigma \to \G$ (for any $x\in \bR$)
 and $\y:\sigma\to \G_+$  by
\be
L(x) := p - \coth(\ad_q) \xi_\sigma - x\,\xi_\sigma,
\qquad
\y:= \y_\M - w^2(\ad_q)\xi_\sigma,
\label{4.31}\ee
equations (\ref{4.29})-(\ref{4.30}) are equivalent to
\be
\dot L(x) = [ \y, L(x)].
\label{4.32} \ee
The conserved quantities associated with this Lax equation
are the same as those exhibited in Theorem 4, since $L(x)$ is
the restriction of the function $K(x)$ (\ref{4.1}) to the gauge slice
$\sigma$,
\be
L(x) = K(x)\vert_\sigma.
\label{4.33}\ee
}

To verify Proposition 5, it is sufficient to
note that $\dot q=p$ and (\ref{4.29})  follow, respectively, from
the $\A$-components of the
first and the second equations under (\ref{4.25}), and (\ref{4.30})
also follows directly from (\ref{4.25}).
(By construction,
the evolution equation just obtained is generated by $\H_{red}$ (\ref{3.25}) through the
Dirac bracket if $\sigma$ is a complete gauge fixing.)
In view of (\ref{4.33}), the conserved quantities in involution described in Theorem 4
receive the usual interpretation as the $G$ invariant functions of the Lax matrix.
This is valid since any point of $P^\cO_{\Psi=0}$ can be
transformed into the gauge slice $\sigma$ by the action of $G_+$, and such a
gauge transformation may be used to convert $K(x)$ into $L(x)$ since $K(x)$ is a $G_+$
equivariant function on $P^\cO$.

If for some  value of $\xi_\sigma$, say $\xi_\sigma = \mu$, with a suitable
function $\y_\M(q,\xi_\sigma)\in \M$,
it so happens that
\be
[ \y_\M(q,\mu) - w^2(\ad_q) \mu, \mu]= 0
\qquad \forall q\in \check \A,
\label{4.35}\ee
then one can `freeze' the spin variable to
that value $\mu$ (see (\ref{4.30})).
By using  the identity
\be
[ w^2(\ad_q) Z, Z] =  (\sinh\ad_q) [  w(\ad_q) Z,  w'(\ad_q) Z]
\qquad \forall Z\in \M^\perp,
\label{4.34}\ee
one can check that (\ref{4.35}) is equivalent to
\be
[ \y_\M(q,\mu), w(\ad_q) \mu] = [w(\ad_q) \mu, w'(\ad_q) \mu ]_{{\cal A}^\perp},
\qquad \forall q\in \check \A,
\label{4.extra}\ee
where the subscript  means projection onto $\A^\perp$ according to (\ref{2.5}).
Equation (\ref{4.extra}) appears\footnote{Equation (\ref{4.extra}) corresponds to (2.22) in \cite{OPInv}
by identifying  $\y_\M$ and $-w(\ad_q)\mu$  with the objects $D$ and
$X$ used there.
In effect,
in \cite{OPInv} an ansatz  was also adopted for $\mu$,
 which is confirmed in the examples of Sect.~5.}
in the work of
Olshanetsky and Perelomov, too,
as the key condition
for obtaining Lax representations for
Calogero type models (\ref{1.1}).
In the cases for which such a constant value $\mu$ exists,
the specializations of our Lax operator $L(x)$ furnished by
\be
L_-=L(0)= p - \coth (\ad_q) \xi_\sigma
\quad\hbox{and}\quad
\cL := e^{-\ad_q} L(1) = p - w(\ad_q) \xi_\sigma,
\label{4.36}\ee
reproduce precisely the alternative Lax operators of \cite{OPInv} upon setting $\xi_\sigma:=\mu$.
Notice that $L(1)$ is essentially the same as the function $L$ introduced
in (\ref{3.21}), and the conjugation by $e^{-q}$ is useful since it leads to a
$\G_-$ valued Lax operator.
By looking at the explicit form of $\H_{red}$ (\ref{3.25}),  it is reasonable
to expect that
(\ref{4.35}) holds only if the reduced Poisson structure of
$\cO_{red}$ vanishes at $[\mu]$.
This is the case automatically whenever $\cO_{red}$ is a trivial
space consisting of a single point,
which is realized in the examples described in the subsequent section.

\section{Spinless models obtainable by the KKS mechanism}
\setcounter{equation}{0}

Here we first recall that
the spinless model (\ref{1.1}) of $A_{k-1}$ type (the `hyperbolic Sutherland model')
arises from the symmetric space $SL(k,\bC)/SU(k)$
by using a minimal coadjoint orbit of $SU(k)$.
In fact,
the reduced orbit (\ref{3.24}) consists of a single point in this case \cite{KKS}.
Relying on the
mechanism that works in this basic example, we
then explain why the spinless $BC_n$ model is associated
with $SU(n+1,n)$, as  presented in \cite{OPInv,OPRep1,Per}  without detailed explanation.

The standard Cartan involution of $G=SL(k,\bC)$  operates as
$\Theta(g) = (g^\dagger)^{-1}$ and
\be sl(k,\bC)= su(k) +\ri\, su(k)
\label{5.1}\ee
is the corresponding Cartan decomposition of the real
simple Lie algebra $\G=sl(k,\bC)$. By using the natural embedding,
we can take $\A= \ri \T_{k-1}$, where $\T_{k-1}$ denotes the
standard Cartan subalgebra of $su(k)$. Then  $\M= \T_{k-1}$ and
$M=\bT_{k-1}$ is the maximal torus of $G_+=SU(k)$. For any $u\in
\bC^k$, viewed as a column vector, we define
\be \eta(u):= \ri
\left( u u^\dagger - \frac{u^\dagger u}{k} {\bf 1}_k\right)\in
su(k), \label{5.2}\ee
with ${\bf 1}_k$ denoting the unit matrix. The
minimal coadjoint orbits of $SU(k)$ are provided by
\be
\cO^{k,\kappa}:= \{ \eta(u) \, \,\vert\, u \in \bC^k,\quad u^\dagger
u = k \kappa \,\}, \label{5.3}\ee
where $\kappa>0$ is a constant.
(Of course, $-{\cal O}^{k,\kappa}$ is also a minimal orbit, but it
either coincides with ${\cal O}^{k,\kappa}$ or is obtained from it
by an automorphism of $su(k)$. Since $\pm {\cal O}^{k,\kappa}$
always lead to similar systems, we may focus on ${\cal O}^{k,\kappa}$.)
We need the constrained orbit
\be
\cO^{k,\kappa}_0:= \{ \eta(u)\in \cO^{k,\kappa}\, \,\vert\,
\eta(u)_{a,a}=0\quad \forall a=1,\ldots, k \,\}. \label{5.4}\ee
Note
that $\cO^{k,\kappa}_0= \cO^{k,\kappa}\cap \M^\perp$ and
 $\eta(u)\in \cO^{k,\kappa}_0$ is associated with $u\in \bC^k$ of the form
\be
u_a = \sqrt{\kappa} e^{\ri \beta_a},
\quad
\beta_a \in \bR, \quad \forall a=1,\ldots, k.
\label{5.5}\ee
This implies
that any $\eta(u) \in \cO^{k,\kappa}_0$
can be transformed by $\bT_{k-1}$ into the representative
$\mu^{k,\kappa}$ furnished by the matrix
\be
(\mu^{k,\kappa})_{a,b} =  \ri  \kappa  ( 1- \delta_{a,b}),
\label{5.6}\ee
showing that
\be
\cO^{k,\kappa}_{red}=
(\cO^{k,\kappa}\cap \M^\perp)/M= \cO^{k,\kappa}_0/\bT_{k-1}
\label{5.7}\ee
consists of a single point indeed.
One can readily calculate that the resulting Hamiltonian is given
by (\ref{1.1}) with $\cR$ now being
the root system of $sl(k,\bC)$ (and $g_\alpha^2 \sim \kappa^2$).
The way whereby the orbital reduced space (\ref{5.7})
is trivial is referred to below as the `KKS mechanism',
since the choice of the orbit (\ref{5.3}) goes
 back to the classical work of
Kazhdan, Kostant and Sternberg \cite{KKS}, where the Sutherland model
was first derived by Hamiltonian reduction.

Now we make the following important observation:
{\em A spinless Calogero model (\ref{1.1})
arises from the symmetric space $G/G_+$ if
the  `KKS mechanism' that works for $SL(k,\bC)$ as described above can be applied by embedding.}
For this  to be realized, $G_+$ must contain a simple factor of
$SU(k)$ type and $M$  must act
on the minimal orbits of this $SU(k)$ factor as the maximal torus
$\bT_{k-1}\subset SU(k)$.
By inspecting the properties of the real simple Lie algebras tabulated in \cite{Knapp},
one sees that {\it these conditions single out the algebras $\G=su(m,n)$, for all $m\geq n$}.
In fact, among the classical Lie algebras  there are no other cases for which $\G_+$
contains $su(k)$
and at the same time $\M$ contains a non-zero Abelian factor.\footnote{This is also true
for
the exceptional Lie algebras apart from $E_6$.
There exists a real form of $E_6$ \cite{Knapp}
for which $\G_+ = su(6) \oplus su(2)$ and
$\M$ is Abelian of dimension 2.
By studying the relative position of $\M$ and the $su(2)$ factor of $\G_+$,
it would be interesting to investigate if the KKS mechanism is applicable in this case or not.}
The system of restricted roots of $su(m,n)$ is of $C_n$ type if $m=n$,
and $BC_n$ type if $m>n$. Nevertheless, as we explain below, the spinless $BC_n$ Calogero
model can only be associated with $su(n+1,n)$.

By using  $I_{m,n}:=\mathrm{diag}(\1_m,  -\1_n)$ with $m\geq n$,
the standard realizations of
the Lie group $SU(m,n)$ and its Lie algebra $su(m,n)$ are
\bea
&& SU(m,n)=\{g\in SL(m+n,\mathbb{C})\,\vert\, g^\dagger I_{m,n}g=I_{m,n}\},
\label{5.8}\\
&& su(m,n)=\{X\in\ sl(m+n,\mathbb{C})\,\vert\, X^\dagger I_{m,n}
+I_{m,n}X=0\}.
\label{5.9}
\eea
Written as a block matrix,  $X\in \G=su(m,n)$ has the form
\be
X=\left(\begin{array}{cc}
A & B\\
B^\dagger & D
\end{array}\right),
\label{5.10}\ee
where $B\in\mathbb{C}^{m\times n}$, $A\in u(m)$, $D\in u(n)$ and $\tr A + \tr D=0$.
The Cartan involution of $G=SU(m,n)$ is
$\Theta: g \mapsto (g^\dagger)^{-1}$, and thus
\be
G_+ = S( U(m) \times U(n)),
\label{5.11}\ee
\be
\G_+= su(m) \oplus su(n) \oplus \bR C_{m,n}=
\left\{\left(\begin{array}{cc}
A & 0\\
0 & D
\end{array}\right) + x C_{m,n}\Bigg|\:
A\in su(m),\, D\in su(n),\, x\in \bR \right\}
\label{5.12}\ee
with the central element
\be
C_{m,n}:=\mathrm{diag}( \ri n \1_m, -\ri m\1_n).
\label{5.13}\ee
A convenient choice for the maximal Abelian subspace of
\be
\G_-=\left\{\left(\begin{array}{cc}
0 & B\\
B^\dagger & 0
\end{array}\right)\Bigg|\:
B\in\mathbb{C}^{m\times n}\right\}
\label{5.14}\ee
is given by
\be
\A:=\left\{
\left(\begin{array}{ccc}
\0_n & 0 & Q\\
0 & \0_{m-n} & 0\\
Q & 0 & \0_n
\end{array}\right)
\in\G_- \:\Bigg|\:
Q=\mathrm{diag}(q^1,\ldots,q^n),\: q^j\in\mathbb{R}\right\}.
\label{5.15}\ee
Taking $\chi:=\mathrm{diag}(\chi_1,\ldots,\chi_n)$ with any $\chi_j\in\mathbb{R}$,
the centralizer of $\A$ in $\G_+$ is
\be
\M=\{\mathrm{diag}(\ri\chi,\gamma,\ri\chi) \: |\:
\gamma\in u(m-n),\,
\tr \gamma + 2\ri \tr\chi =0\},
\label{5.16}\ee
and the corresponding subgroup of $G_+$ is
\be
M=\{ \diag( e^{\ri \chi}, \Gamma, e^{\ri\chi}) \: | \:   \Gamma \in U(m-n),\,
(\det \Gamma)(\det e^{\ri 2\chi}) =1 \}.
\label{5.17}\ee

Let $\cO^k$ be an arbitrary orbit of $SU(k)$. For $k\in \{m,n\}$, denote
by $\tilde \cO^k$ the natural embedding of $\cO^k$ into an $su(k)$ factor
of $\G_+$ (\ref{5.12}).
If $m\neq n$, then the most general coadjoint orbit of $G_+$ (\ref{5.11}) has the form
\be
\cO= \tilde \cO^m + \tilde \cO^n + x C_{m,n}\qquad (x\in  \bR).
\label{5.18} \ee
In the $SU(n,n)$ case we can similarly
embed different orbits of $SU(n)$ into the two isomorphic factors.
In order to get a trivial reduced space $\cO\cap \M^\perp/M$ by the KKS mechanism,
we must take the constituent orbits to be of the type $\cO^{k,\kappa}$ (\ref{5.3}).
If $m \leq (n+1)$, then a simple dimension counting argument says that
$\cO\cap \M^\perp/M$ could possibly  be a trivial space only if
either $\tilde\cO^m$ or $\tilde\cO^n$ in (\ref{5.18}) is taken to be zero.
Furthermore, it follows from the structure of $\M$ (\ref{5.16}) that
if $m>(n+1)$, then the KKS mechanism could be applicable only if the non-trivial
component of $\cO$  is contained in the factor of size $n$.
Detailed, simple inspection leads to the following result.

\bigskip
\noindent
{\bf Theorem 6.}
{\em
For any $m>n$, and similarly for $m=n$,
consider the family (\ref{5.18}) of non-zero  coadjoint orbits of $SU(m,n)$
with minimal orbits (\ref{5.3}) as constituents.
In this family,
\be
\cO \cap \M^\perp/M
\label{5.19}\ee
consists of a single point precisely in the following cases:
\be
\hbox{for any $m\geq n$, the orbits of type}\quad  \tilde \cO^{n, \kappa}\quad
\forall \kappa >0,
\label{5.20}\ee
\be
\hbox{for $m=n$, all non-zero orbits of the form}\quad  \tilde \cO^{n, \kappa}+  x C_{n,n}
\quad  \forall\kappa\geq 0,\, x\in \bR,
\label{5.21}\ee
\be
\hbox{for $m=(n+1)$, the orbits}\quad  \tilde \cO^{n+1, \kappa}+  x C_{n+1,n} \quad
\forall \kappa> 0,\, x\in \bR \quad \hbox{satisfying}\quad
\label{5.22}\ee
\qquad\,\, in this case  $(\kappa-nx)\geq 0$ and $(\kappa+x) \geq 0$.}

\medskip\noindent
{\bf Proof.}
Let us consider an orbit of $SU(n+1,n)$ of the form
\be
\cO:=
\tilde \cO^{n+1, \kappa}+  x C_{n+1,n}.
\label{5.23}\ee
The general element $\xi \in \cO$ can be written as
\be
\xi = \tilde \eta + x C_{n+1,n},
\label{5.24}\ee
where $\tilde \eta\in \tilde\cO^{n+1,\kappa}$ is the
embedding of $\eta \in \cO^{n+1,\kappa}$ into the factor $su(n+1)$ of $\G_+$ (\ref{5.12}).
By the mapping $\eta \mapsto \xi$ according to (\ref{5.24}),
the constraint
\be
\tr (X \xi)=0\qquad \forall X\in \M
\label{5.25}\ee
turns out to be equivalent to
\be
\eta_{diag}= \diag( \ri x \1_n, -\ri xn ),
\label{5.26}\ee
where $\eta_{diag}$ denotes the diagonal part of the matrix $\eta \in su(n+1)$.
Thus we have a  one-to-one correspondence between
 $\cO \cap \M^\perp$ and the `constrained KKS orbit'
 $\cO^{n+1,\kappa}_x$ consisting of the elements $\eta\in \cO^{n+1,\kappa}$
 subject to (\ref{5.26}).
By this correspondence, the action of $M$ (\ref{5.17}) on
$\cO\cap \M^\perp$ can be represented as the action of $\bT_{n}\subset SU(n+1)$ on $\cO^{n+1,\kappa}_x$,
which gives rise to a one-to-one map
\be
\cO\cap \M^\perp/M  \longleftrightarrow
\cO^{n+1,\kappa}_x/\bT_n.
\label{5.27}\ee
Therefore we have to show that the latter space consists of a single point.
Now write any $\eta \in \cO^{n+1,\kappa}$ as $\eta(u)$ with
\be
u= (u_1, \ldots, u_n, u_{n+1})^t,
\label{5.28}\ee
using the notation (\ref{5.2}).
The constraint (\ref{5.26}) requires that
\be
\vert u_j\vert^2 = \left( \kappa + x\right)\quad \forall j=1,\ldots, n,
\quad\hbox{and}\quad
\vert u_{n+1} \vert^2 = \left( \kappa - xn\right).
\label{5.29}\ee
Hence the constants $\kappa$ and $x$ has to be chosen so that the right hand sides above
are non-negative.
Now the point is that given the constraint (\ref{5.29}), we can bring any $u$ by a
$\bT_n$ transformation to the following normal form, say $\hat u$:
\be
 \hat u_j = e^{\ri \alpha}\sqrt{ \kappa + x }\quad \forall j=1,\ldots, n,
 \qquad
 \hat u_{n+1} =e^{\ri\alpha} \sqrt{ \kappa - xn}
\label{5.30}\ee
with some $\alpha\in \bR$.
Since $u$ matters only up to phase,
$\eta( e^{i\alpha} u)= \eta(u)$,
we see
 that every point of
$\cO^{n+1,\kappa}_x$ can be transformed into
$\eta(\hat u)$ by the action of $\bT_n$.
This means that $\cO^{n+1,\kappa}_x$ consists of a single orbit of $\bT_n$, and may be
represented by $\eta(\hat u)$.
By the correspondence (\ref{5.27}), it follows that the reduction of $\cO$ (\ref{5.23})
 by $M$ yields a
trivial space, and as a representative of this space one may take the matrix
\be
\xi_{red}:= \tilde \eta(\hat u) + x C_{n+1,n} \in su(n+1,n).
\label{5.31}\ee
It is quite similar but even simpler to verify that (\ref{5.19})
consists of a single element also
in the cases listed under (\ref{5.20}) and (\ref{5.21}).
In all other cases when simple counting
does not exclude that $\cO\cap \M^\perp$ contains a single orbit of $M$,
the constraints, $\tr(X\xi)=0$  $\forall X\in \M$, are found to be inconsistent with
the form of the orbits considered.
For example, one can  check for $SU(n+1,n)$ that
\be
(\tilde \cO^{n, \kappa} + x C_{n+1,n})\cap \M^\perp=\emptyset
\qquad
\forall x\neq 0,\,\kappa\geq 0.
\label{5.32}\ee
Hence we may conclude that
the list given in the theorem exhausts all cases for which
(\ref{5.19}) consists of a single point by the KKS mechanism.
\emph{Q.E.D.}

\medskip

One may use the conventions collected in the appendix to obtain the
Hamiltonians of the spinless Calogero models corresponding to the various
cases listed under Theorem 6.
In the most complicated case of equation (\ref{5.22}),
it is easily checked that the representative $\xi_{red}$ (\ref{5.31}) of the
reduced orbit can be expanded in the form
\be
\xi_{red}=
 2g  \sum_{1\leq k<l\leq n}(E^{+,\ri}_{e_k-e_l}+E^{+,\ri}_{e_k+e_l})
+ 2g_1\sum_{k=1}^n E^{+,\ri}_{e_k}
+2g_2\sum_{k=1}^n E^{+,\ri}_{2e_k} ,
\label{5.33}\ee
where we use the basis introduced in (\ref{A.6})-({\ref{A.13}) together with
the notation
\be
g:=\frac{\kappa +x}{2},\qquad
g_1:=\sqrt{\frac{(\kappa +x)(\kappa -nx)}{2}},\qquad
g_2:=\frac{(n+1)x}{\sqrt{2}}.
\label{5.34}\ee
 From (\ref{4.36}) with (\ref{4.28}), the corresponding Lax operator is
\be
\cL(q,p)= p - w(\ad_q) \xi_{red},
\label{5.35}\ee
where $q\in \check \A$ is parametrized by $\mathrm{diag}(q^1,\ldots,q^n)$
according to (\ref{5.15}), now with $m=n+1$, and $p\in \A$
is similarly parametrized by
$\mathrm{diag}(p_1,\ldots,p_n)$.
This leads to the Hamiltonian
\bea
&& H_{BC_n}(q,p):= \frac{1}{4} \tr (\cL(q,p))^2 =\frac{1}{2}\sum_{k=1}^n p_k^2
+ \sum_{k=1}^n\frac{g_1^2}{\sinh^2(q^k)}
+\sum_{k=1}^n\frac{g_2^2}{\sinh^2(2q^k)}
\nonumber\\
&& \qquad\qquad \qquad  + \sum_{1\leq k<l\leq n}\frac{g^2}{\sinh^2(q^k-q^l)}
+\sum_{1\leq k<l\leq n}\frac{g^2}{\sinh^2(q^k+q^l)}\,.
\label{5.36}\eea
On account of (\ref{5.34}), the coupling constants satisfy the quadratic relation
\be
g_1^2 -2g^2 +\sqrt{2} gg_2=0.
\label{5.37}\ee
One can similarly spell out the Hamiltonian in the other cases of Theorem 6.
Although we do not obtain any
spinless models that were not described before in the symmetric space framework,
it is worth summarizing the list of the resulting models as a proposition.

\bigskip
\noindent
{\bf Proposition 7.}
{\em
The Calogero type Hamiltonian corresponding to case (\ref{5.22}) of Theorem 6
is $H_{BC_n}$ (\ref{5.36}) with the relation (\ref{5.37}).
The Hamiltonian in the case (\ref{5.21}) turns out to be
\be
H_{C_n}(q,p)=\frac{1}{2} \sum_{k=1}^n p_k^2
+\sum_{1\leq k<l\leq n}\frac{\kappa^2/4}{\sinh^2(q^k-q^l)}
+\sum_{1\leq k<l\leq n}\frac{\kappa^2/4}{\sinh^2(q^k+q^l)}
+\sum_{k=1}^n\frac{n^2 x^2/2}{\sinh^2(2q^k)}\,.
\ee
The orbit (\ref{5.20}) leads to
\be
H_{D_n}(q,p)=\frac{1}{2} \sum_{k=1}^n p_k^2
+\sum_{1\leq k<l\leq n}\frac{\kappa^2/4}{\sinh^2(q^k-q^l)}
+\sum_{1\leq k<l\leq n}\frac{\kappa^2/4}{\sinh^2(q^k+q^l)}\,.
\ee
}

\bigskip

The statement of Proposition 7 amounts to a systematization of known results.
Indeed,
the Lax matrix (\ref{5.35}) of the $BC_n$ model (\ref{5.36})
reproduces\footnote{Our conventions are chosen so that
(\ref{5.36}) reproduces the $BC_n$ Hamiltonian as given
by Olshanetsky and Perelomov and (\ref{5.37})  coincides with
the correct relation (B.11) in \cite{OPInv}.
(This quadratic relation is mistyped in (3.3) in \cite{OPInv} and also
in \cite{OPRep1,Per} where $g_1 g_2$ appears in place of  $gg_2$.)
Our Lax pair, defined by (\ref{5.35}) with Proposition 5 and (\ref{4.36}),
reproduces their $BC_n$ Lax pair after
similarity transformation by a constant matrix.}
the original result of Olshanetsky and Perelomov.
The $C_n$ and $D_n$  models obtained by the KKS mechanism are essentially degenerations
of the $BC_n$ model.
The $C_n$ model is treated in \cite{ABT}
 by using Hamiltonian reduction (see also \cite{FW}).
The constants of motion provided by the
eigenvalues of the Lax matrix (\ref{5.35}) guarantee Liouville integrability
if (\ref{5.37}) holds, but one needs a different approach for proving
that the $BC_n$ model (\ref{5.36})
is also integrable with three arbitrary coupling constants \cite{Sas2}.

\medskip
\noindent
{\bf Remark 8.}
For $\G:=su(m,n)$ with any $m\geq n$,
let us consider the functions
\be
f_k: \G\to \bR,
\qquad
f_k(X):= \tr ((ABDB^\dagger)^k)\qquad (k=1,\ldots, n),
\label{5.40}\ee
where $X\in \G$ is written in the form (\ref{5.10}).
These functions are $G_+=S(U(m) \times U(n))$ invariant,
and hence give rise to conserved quantities for all spin Calogero models
based on $su(m,n)$ as follows from Theorem 4.
For the models provided by Proposition 7 these constants of motion
are not independent from the eigenvalues of the Lax matrices in (\ref{4.36}).
This can be seen  by combining Theorem 4 with the fact \cite{Per} that the eigenvalues
of any of the two  Lax matrices (\ref{4.36})  generate the same
{\em maximal set of constants of motion in involution}
for the above spinless models.
However,
for general spin Calogero models based on $su(m,n)$ the
conserved quantities associated with the functions $f_k$ are
independent from the conserved quantities in involution furnished by
Theorem 4.
In fact, we have checked in several cases (even numerically) that the functions
$f_k \circ K(x)$ Poisson commute neither with each other for different $k$  nor
with all of the invariants in involution $h\circ K(y)$ ($\forall h\in C^\infty_G(\G)$)
described in Theorem 4.
It could be an interesting problem for the future to clarify the various possible
(Liouville, degenerate, super)
integrability properties of the spin Calogero models that we obtained, and in
particular to understand the role of the conserved quantities just exhibited.

\section{Discussion}
\setcounter{equation}{0}

In this paper we investigated the symmetry reductions of the geodesic motion on
a symmetric space of negative curvature $G/G_+$ based on the
action of $G_+$ on $G/G_+$.
Taking an \emph{arbitrary} value of the momentum map
and restricting to regular elements in the configuration space, the
result turned out to be the hyperbolic spin Calogero model characterized by Theorem 1.
We analyzed the integrability properties of this family of models in
Section 4, describing  many conserved quantities in Theorem 4 and a spectral
parameter dependent Lax pair in Proposition 5.
In Section 5
we classified the cases yielding spinless Calogero models of type (\ref{1.1})
relying on the KKS mechanism.
We conjecture that no other spinless models arise in the Hamiltonian reduction
framework, even without assuming the applicability of the KKS mechanism.

Trigonometric spin Calogero models  appear similarly
in the positive curvature case,
the corresponding rational models are related to the symmetric spaces of zero
curvature \cite{AKLM,Hoch}, and analogous elliptic  models
should also exist.
Further generalizations can be obtained,
for instance,  by reducing the geodesic motion
on affine symmetric spaces \cite{OshSek}.
In fact, this yields
spin extensions of the Calogero models attached to root systems
with signature \cite{OR,Hashi}.
All these examples fit in the theory of singular symplectic reduction
of cotangent bundles with a single isotropy type
in the configuration space (\cite{OrRat,Hoch,Hoch2} and references therein).

Let us recall \cite{BV} that,
in classical integrable systems that admit a diagonalizable Lax matrix with Poisson commuting
eigenvalues, the Poisson brackets between the matrix
elements of the Lax matrix are always encoded by some classical $r$-matrix
that may depend on the dynamical variables.
Starting from \cite{AT}, a lot of effort went into finding the dynamical $r$-matrices
of Calogero type models.
In the Hamiltonian reduction setting the integrability properties can be analyzed directly,
but we are nevertheless interested in the corresponding dynamical $r$-matrices, too.
So far we computed the $r$-matrix belonging to
the Lax matrix $\cL$ (\ref{4.36}) by using a complete (local)
gauge fixing $\sigma \subset S$ of  type (\ref{4.24}).
With the usual St Petersburg notation,
we found that the Dirac brackets associated with the gauge fixing
can be written as follows:
\be
\{ \cL_1,\cL_2\}^*=  [ r_{12} + d_{12}, \cL_1] - [ r_{21} + d_{21}, \cL_2],
\label{6.1}\ee
where $r_{12}\in \M^\perp \otimes \A^\perp$ depends on the variable $q$ as
\be
r_{12}(q)= \sum_{\alpha\in \cR_+} \sum_{k=1}^{\nu_\alpha} \coth\alpha(q)
E_\alpha^{+,k}\otimes E_\alpha^{-,k}
\label{6.2}\ee
and $d_{12}\in \M\otimes \A^\perp$ depends in general also on the spin variable as
\be
d_{12}(q, \xi_\sigma) = \sum_{\alpha, k, b} D^\alpha_{k,b}(\xi_\sigma) (\sinh \alpha(q))^{-1}
M^b \otimes E_\alpha^{-,k}
\label{6.3}\ee
with coefficients $D^\alpha_{k,b}(\xi_\sigma)$ determined  by the constraints
defining the gauge fixing.
Here $\{ M^b\}$ denotes a basis of $\M$, and  $\xi_\sigma$ becomes a
constant if $\cO_{red}$ consists of one point.
In the spinless examples listed in Section 5 equation (\ref{6.1})  reproduces and extends previous
results of \cite{ABT,FW}.
Details will be presented elsewhere.

Finally, let us briefly discuss the quantization of the models (\ref{1.2}).
Clearly,
the quantum mechanical analogue of the phase space (\ref{3.6}) is
$L^2(G/G_+, V_\Lambda)$, i.e., $V_\Lambda$ valued square-integrable functions
on $G/G_+$. Here $V_\Lambda$ is an irreducible representation of $G_+$ corresponding to
a quantizable coadjoint orbit $\cO$.
It is easy to see that quantum Hamiltonian reduction requires restriction to the $G_+$ equivariant
wave functions in the Hilbert space, which  can be represented by
functions on the Weyl chamber $\check \A$ with values in $V_\Lambda[0]$, where
$V_\Lambda[0]$ consists of
the invariants  in $V$ with respect to the action of the subgroup $M\subset G_+$.
The Casimirs of $\G$ yield
commuting self-adjoint operators on the reduced Hilbert space
formed by the these $V_\Lambda[0]$ valued functions.
Thus the quadratic Casimir gives rise to the Hamiltonian of the spin Calogero model,
which is a spinless model at the quantum mechanical level if and only if
$\mathrm{dim}(V_\Lambda[0])=1$.
The analysis of quantum (spin) Calogero models translates in this way into problems
in harmonic analysis and representation theory.
Quite an analogous procedure can be applied starting with
 positive or zero curvature symmetric spaces, too.
We plan to elaborate the quantization in the future
building on the previous works dealing with special cases \cite{OPRep2}.

\bigskip
\bigskip
\noindent{\bf Acknowledgements.}
The work of L.F. was supported in part by the Hungarian
Scientific Research Fund (OTKA) under the grants
T043159, T049495  and by the EU networks `EUCLID'
(contract number HPRN-CT-2002-00325) and `ENIGMA'
(contract number MRTN-CT-2004-5652).
He is indebted to G. Felder, S. Hochgerner and T. Ratiu  for useful discussions
and  to J. Balog for comments on the manuscript.
B.G.P. is grateful for support by a CRM-Concordia Postdoctoral
Fellowship and he especially  wishes to thank J. Harnad for
hospitality in Montreal.

\renewcommand{\theequation}{\arabic{section}.\arabic{equation}}
\renewcommand{\thesection}{\Alph{section}}
\setcounter{section}{0}

\section{Restricted roots and convenient basis for $su(m,n)$}
\setcounter{equation}{0}
\renewcommand{\theequation}{A.\arabic{equation}}

For reference in the main text, in this appendix we present the restricted roots
and the basis elements $E_\alpha^{\pm, k}$ (\ref{2.6}) explicitly
for $\G:=su(m,n)$ with any $m\geq n$,  using
the realization of this real simple Lie algebra and its Cartan involution
specified in (\ref{5.8})-(\ref{5.15}).

Now it proves convenient to present any matrix $X\in su(m,n)$ in a block-form
corresponding to the partition $(m+n) = n + (m-n) +n$, i.e.,
\be
X=\left(\begin{array}{ccc}
a & v & b\\
-v^\dagger & e & w\\
b^\dagger & w^\dagger & d
\end{array}\right),
\qquad \tr a + \tr e + \tr d=0,
\label{A.1}\ee
where $a,d\in u(n)$, $e\in u(m-n)$ and $v\in\mathbb{C}^{n\times (m-n)}$
parametrize $\G_+$,
and
$b\in\mathbb{C}^{n\times n}$, $w\in\mathbb{C}^{(m-n)\times n}$
parametrize $\G_-$.
Writing the general element of $\A$ (\ref{5.15}) as
\be
q:=
\left(\begin{array}{ccc}
0 & 0 & Q\\
0 & 0 & 0\\
Q & 0 & 0
\end{array}\right)
\quad\hbox{with}\quad
Q=\mathrm{diag}(q^1,\ldots,q^n),\quad q^j\in\mathbb{R},
\label{A.2}\ee
one may introduce the functionals $e_k\in \A^*$ ($k=1,\ldots, n$) by
$e_k(q):= q^k$.
The system of restricted roots, $\cR$, is of $BC_n$ type if $m>n$ and of
$C_n$ type if $m=n$.
Indeed, $\cR$
is given by $\cR=\cR _+ \cup (-\cR_+)$ with
\be
\cR_+:=\{e_k\pm e_l\:(1\leq k<l\leq n),\, 2e_k,  e_k\:(1\leq k\leq n)\}
\quad \hbox{if}\quad m>n,
\label{A.3}\ee
and
\be
\cR_+:=\{e_k\pm e_l\:(1\leq k<l\leq n),\, 2e_k \:(1\leq k\leq n)\}
\quad \hbox{if}\quad m=n.
\label{A.4}\ee
The corresponding multiplicities are
\be
\nu_{e_k \pm e_l}= 2 \quad (1\leq k<l\leq n),\quad
\nu_{2e_k}= 1\quad\hbox{and}\quad
\nu_{e_k}= 2(m-n) \quad (1\leq k\leq n).
\label{A.5}\ee
Instead of the restricted root vectors $E_\alpha^j$
for which $[q, E_\alpha^j] = \alpha(q) E_\alpha^j$,
we directly list their linear combinations (\ref{2.6}) lying in $\G_\pm$.
The two-dimensional subspaces of $\M^\perp \subset \G_+$ associated
with $(e_k \pm e_l)\in \cR_+$  for any $1\leq k<l\leq n$
are spanned by the matrices
\be
E^{+,\rr}_{e_k\pm e_l}:=\frac{1}{2}\left(
\begin{array}{ccc}
E_{kl}-E_{lk} & 0 & 0\\
0 & 0 & 0\\
0 & 0 & \mp (E_{kl}-E_{lk})
\end{array}
\right),
\label{A.6}\ee
and
\be
E^{+,\ri}_{e_k\pm e_l}:=\frac{\ri}{2}\left(
\begin{array}{ccc}
E_{kl}+E_{lk} & 0 & 0\\
0 & 0 & 0\\
0 & 0 & \mp (E_{kl}+E_{lk})
\end{array}\right),
\label{A.7}\ee
whose real or imaginary character is indicated by the superscripts
$\rr$ or $\ri$, respectively.
The generators corresponding to $2e_k\in \cR_+$ are the imaginary matrices
\be
E^{+,\ri}_{2e_k}:= \frac{\ri}{\sqrt{2}}\left(
\begin{array}{ccc}
 E_{kk} & 0 & 0\\
0 & 0 & 0\\
0 & 0 & - E_{kk}
\end{array}
\right).
\label{A.8}\ee
If $m>n$, then the
$2(m-n)$ basis vectors of the subspace of $\G_+$ belonging to
$e_k\in \cR_+$ are
\be
E^{+,\rr,d}_{e_k}:=\frac{1}{\sqrt{2}}\left(
\begin{array}{ccc}
0 & \ E_{kd} & 0\\
- E_{dk} & 0 & 0\\
0 & 0 & 0
\end{array}
\right) \quad\hbox{and}\quad
 E^{+,\ri,d}_{e_k} :=
\frac{\ri}{\sqrt{2}} \left(
\begin{array}{ccc}
0 &  E_{kd} & 0\\
 E_{dk} & 0 & 0\\
0 & 0 & 0
\end{array}\right)
\label{A.9}\ee
for $1\leq d\leq m-n$.
Similarly, the basis of $\A^\perp \subset \G_-$ is given
by the matrices
\be
E^{-,\rr}_{e_k\pm e_l}:= \frac{1}{2}\left(
\begin{array}{ccc}
0 & 0 & E_{lk}\mp E_{kl}\\
0 & 0 & 0\\
E_{kl}\mp E_{lk} & 0 & 0
\end{array}
\right)
\label{A.10}\ee
and
\be
 E^{-,\ri}_{e_k\pm e_l}:=
\frac{\ri}{2}\left(
\begin{array}{ccc}
0 & 0 & - (E_{lk}\pm E_{kl})\\
0 & 0 & 0\\
E_{kl}\pm E_{lk} & 0 & 0
\end{array}
\right), \quad 1\leq k<l\leq n,
\label{A.11}\ee
together with
\be
 E^{-,\ri}_{2e_k}:=
\frac{\ri}{\sqrt{2}}\left(
\begin{array}{ccc}
0 & 0 & - E_{kk}\\
0 & 0 & 0\\
E_{kk} & 0 & 0
\end{array}
\right),
\quad 1\leq k\leq n,
\label{A.12}\ee
and
\be
E^{-,\rr,d}_{e_k}:=
\frac{1}{\sqrt{2}} \left(
\begin{array}{ccc}
0 & 0 & 0\\
0 & 0 & E_{dk}\\
0 &  E_{kd} & 0
\end{array}
\right),
\qquad
 E^{-,\ri,d}_{e_k}:=\frac{\ri}{\sqrt{2}} \left(
\begin{array}{ccc}
0 & 0 & 0\\
0 & 0 & - E_{dk}\\
0 &  E_{kd} & 0
\end{array}
\right)
\label{A.13}\ee
for any $1\leq k\leq n$, $1\leq d\leq m-n$.
Combined with bases of $\A$ and $\M$, the matrices listed under (\ref{A.6})-(\ref{A.13})
span $su(m,n)$.
Their normalization is fixed according (\ref{2.4}), (\ref{2.6})  with
 $\langle X,Y\rangle := \tr(XY)$.
If desired,  the restricted root vectors $E_{\pm\alpha}^j$  can be
recovered easily since \be [q, E_\alpha^{\pm,j}]= \alpha(q)
E_\alpha^{\mp,j}\quad \forall \alpha\in \cR_+,\, j=1,\ldots,
\nu_\alpha. \label{A.14}\ee
In the above formulae of the basis elements
the $E_{kl}$ and so on stand for
the usual elementary matrices of suitable size given according to (\ref{A.1}).

\end{document}